\documentclass[sigconf,natbib=false]{acmart}
\usepackage{listings}
\usepackage{enumerate} 
\usepackage{mdframed}

\newsavebox{\LstBox}

\makeatletter
\renewcommand\@formatdoi[1]{\ignorespaces}
\acmISBN{}
\makeatother
\setcopyright{iw3c2w3}

\title{Looking Back at Postgres}

\author{Joseph M. Hellerstein}
\email{hellerstein@berkeley.edu}

\begin{abstract}
This is a recollection of the UC Berkeley Postgres project, which was led by Mike Stonebraker from the mid-1980's to the mid-1990's.  The article was solicited for Stonebraker's Turing Award book~\cite{stonebrakerbook}, as one of many personal/historical recollections. As a result it focuses on Stonebraker's design ideas and leadership. But Stonebraker was never a coder, and he stayed out of the way of his development team. The Postgres codebase was the work of a team of brilliant students and the occasional university ``staff programmers'' who had little more experience (and only slightly more compensation) than the students. I was lucky to join that team as a student during the latter years of the project. I got helpful input on this writeup from some of the more senior students on the project, but any errors or omissions are mine. If you spot any such, please contact me and I will try to fix them.
\end{abstract}

\begin{document}
\maketitle

\section{Opening}
Postgres was Michael Stonebraker's most ambitious project---his grand effort to build a one-size-fits-all database system. A decade long, it generated more papers, Ph.D.s, professors, and companies than anything else he did. It also covered more technical ground than any other single system he built. Despite the risk inherent in taking on that scope, Postgres also became the most successful software artifact to come out of Stonebraker's research groups, and his main contribution to open source. It is an example of a ``second system''~\cite{brooks1975mythical} that succeeded. As of the time of writing--over thirty years since the project started---the open-source PostgreSQL system is the most popular independent open-source database system in the world, and the fourth most popular database system in the world. Meanwhile, companies built from a Postgres base have generated a total of over \$2.6 billion in acquisitions. By any measure, Stonebraker's Postgres vision resulted in enormous and ongoing impact.

\subsection{Context}

Stonebraker had enormous success in his early career with the Ingres research project at Berkeley~\cite{stonebraker1976design}, and the subsequent start-up he founded with Larry Rowe and Eugene Wong: Relational Technology, Inc. (RTI).

As RTI was developing in the early 1980s, Stonebraker began working on database support for data types beyond the traditional rows and columns of Codd's original relational model. A motivating example current at the time was to provide database support for Computer-Aided Design (CAD) tools for the microelectronics industry. In a paper in 1983, Stonebraker and students Brad Rubenstein and Antonin Guttman explained how that industry needed support for "new data types such as polygons, rectangles, text strings, etc." "efficient spatial searching" "complex integrity constraints" and "design hierarchies and multiple representations" of the same physical constructions~\cite{stonebraker1983application}. Based on motivations such as these, the group started work on indexing (including Guttman's influential R-trees for spatial indexing~\cite{Guttman:1984:RDI:602259.602266}, and on adding Abstract Data Types (ADTs) to a relational database system. ADTs were a popular new programming language construct at the time, pioneered by subsequent Turing Award winner Barbara Liskov and explored in database application programming by Stonebraker's new collaborator, Larry Rowe. In a paper in SIGMOD Record in 1983~\cite{ong1983implementation}, Stonebraker and students James Ong and Dennis Fogg describe an exploration of this idea as an extension to Ingres called ADT-Ingres, which included many of the representational ideas that were explored more deeply---and with more system support---in Postgres.

\section{Postgres: An Overview}

As indicated by the name, Postgres was "Post-Ingres": a system designed to take what Ingres could do, and go beyond. The signature theme of Postgres was the introduction of what he eventually called \textit{Object-Relational} database features: support for object-oriented programming ideas within the data model and declarative query language of a database system. But Stonebraker also decided to pursue a number of other technical challenges in Postgres that were independent of object-oriented support, including active database rules, versioned data, tertiary storage, and parallelism. 

Two papers were written on the design of Postgres: an early design in SIGMOD 1986~\cite{Stonebraker:1986:DP:16894.16888} and a "mid-flight" design description in CACM 1991~\cite{Stonebraker:1991:PNG:125223.125262}. The Postgres research project ramped down in 1992 with the founding of Stonebraker's Illustra startup, which involved Stonebraker, key Ph.D. student Wei Hong, and then-chief-programmer Jeff Meredith. In Figure~\ref{fig:features}, the features mentioned in the 1986 paper are marked with an asterisk*; those from the 1991 paper that \textit{were not} in the 1986 paper are marked with a $\mbox{dagger}^\dagger$. Other goals listed below were tackled in the system and the research literature, but not in either design paper.
Many of these topics were addressed in Postgres well before they were studied or reinvented by others; in many cases Postgres was too far ahead of its time and the ideas caught fire later, with a contemporary twist.

\begin{figure}[t]
\begin{mdframed}
\small
\begin{enumerate}[1. ]

\item Supporting ADTs in a Database System

\begin{enumerate}[a.]

\item Complex Objects (i.e., nested or non-first-normal form data)*

\item User-Defined Abstract Data Types and Functions*

\item Extensible Access Methods for New Data Types*

\item Optimizer Handling of Queries with Expensive UDFs

\end{enumerate}
\item Active Databases and Rules Systems (Triggers, Alerts)*

\begin{enumerate}[a.]

\item Rules implemented as query $\mbox{rewrites}^\dagger$

\item Rules implemented as record-level $\mbox{triggers}^\dagger$

\end{enumerate}
\item Log-centric Storage and Recovery

\begin{enumerate}[a.]

\item Reduced-complexity recovery code by treating the log as data,* using non-volatile memory for commit $\mbox{status}^\dagger$

\item No-overwrite storage and time travel $\mbox{queries}^\dagger$

\end{enumerate}
\item Support for querying data on new deep storage technologies, notably optical disks*

\item Support for multiprocessors or custom processors*

\item Support for a variety of language models

\begin{enumerate}[a.]

\item Minimal changes to the relational model and support for declarative queries*

\item Exposure of "fast path" access to internal APIs, bypassing the query $\mbox{language}^\dagger$

\item Multi-lingual $\mbox{support}^\dagger$

\end{enumerate}
\end{enumerate}
\end{mdframed}
\caption{Postgres features first mentioned in the 1986 paper* and the 1991 $\mbox{paper}^\dagger$.}
\label{fig:features}
\end{figure}

We briefly discuss each of these Postgres contributions, and connections to subsequent work in computing.

\subsection{Supporting ADTs in a Database System}

The signature goal of Postgres was to support new Object-Relational features: the extension of database technology to support a combination of the benefits of relational query processing and object-oriented programming. Over time the Object-Relational ideas pioneered in Postgres have become standard features in most modern database systems.

\subsubsection{Complex objects}

It is quite common for data to be represented in the form of nested bundles or ``objects.'' A classic example is a purchase order, which has a nested set of products, quantities, and prices in the order. Relational modeling religion dictated that such data should be restructured and stored in an unnested format, using multiple flat entity tables (orders, products) with flat relationship tables (product\_in\_order) connecting them. The classic reason for this flattening is that it reduces duplication of data (a product being described redundantly in many purchase orders), which in turn avoids complexity or errors in updating all redundant copies. But in some cases you want to store the nested representation, because it is natural for the application (say, a circuit layout engine in a CAD tool), and updates are rare. This data modeling debate is at least as old as the relational model.

A key aspect of Postgres was to "have your cake and eat it too" from a data modeling perspective: Postgres retained tables as its "outermost" data type, but allowed columns to have "complex" types including nested tuples or tables. One of its more esoteric implementations, first explored in the ADT-Ingres prototype, was to allow a table-typed column to be specified declaratively as a query definition: "Quel as a data type"~\cite{Stonebraker:1984:QDT:602259.602287}.

The ``post-relational'' theme of supporting both declarative queries and nested data has recurred over the years---often as an outcome of arguments about which is better. At the time of Postgres in the 1980s and 1990s, some of the object-oriented database groups picked up the idea and pursued it to a standard language called OQL, which has since fallen from use. 

Around the turn of the millennium, declarative queries over nested objects became a research obsession for a segment of the database community in the guise of XML databases; the resulting XQuery language (headed by Don Chamberlin of SQL fame) owes a debt to the complex object support in Postgres' Postquel language. XQuery had broad adoption and implementation in industry, but never caught on with users. The ideas are being revisited yet again today in query language designs for the JSON data model popular in browser-based applications. Like OQL, these languages are in many cases an afterthought in groups that originally rejected declarative queries in favor of developer-centric programming (the "NoSQL" movement), only to want to add queries back to the systems post-hoc. In the meantime, as Postgres has grown over the years (and shifted syntax from Postquel to versions of SQL that reflect many of these goals), it has incorporated support for nested data like XML and JSON into a general-purpose DBMS without requiring any significant rearchitecting. The battle swings back and forth, but the Postgres approach of extending the relational framework with extensions for nested data has shown time and again to be a natural end-state for all parties after the arguments subside.

\subsubsection{User-defined abstract data types and functions}

In addition to offering nested types, Postgres pioneered the idea of having opaque, extensible Abstract Data Types (ADTs), which are stored in the database but not interpreted by the core database system. In principle this was always part of Codd's relational model: integers and strings were traditional, but really any atomic data types with predicates can be captured in the relational model. The challenge was to provide that mathematical flexibility in software. To enable queries that interpret and manipulate these objects, an application programmer needs to be able to register User-Defined Functions (UDFs) for these types with the system, and be able to invoke those UDFs in queries. User-Defined Aggregate (UDA) functions are also desirable to summarize collections of these objects in queries. Postgres was the pioneering database system supporting these features in a comprehensive way.

Why put this functionality into the DBMS, rather than the applications above? The classic answer was the significant performance benefit of ``pushing code to data,'' rather than ``pulling data to code.'' Postgres showed that this is quite natural within a relational framework: it involved modest changes to a relational metadata catalog, and mechanisms to invoke foreign code, but the query syntax, semantics, and system architecture all worked out simply and elegantly.

Postgres was a bit ahead of its time in exploring this feature. In particular, the security implications of uploading unsafe code to a server were not an active concern in the database research community at the time. This became problematic when the technology started to get noticed in industry. Stonebraker commercialized Postgres in his Illustra start-up, which was acquired by Informix in large part for its ability to support extensible ``DataBlades'' (extension packages) including UDFs. Informix's Postgres-based technology, combined with their strong parallel database offering, made Informix a significant threat to Oracle. Oracle invested heavily in negative marketing about the risks of Informix's ability to run ``unprotected'' user-defined C code. Some trace the demise of Informix to this campaign, although Informix's financial shenanigans (and subsequent federal indictment of its then-CEO) were certainly more problematic. Now, decades later, all the major database vendors support the execution of user-defined functions in one or more languages, using newer technologies to protect against server crashes or data corruption.

Meanwhile, the Big Data stacks of the 2000s---including the MapReduce phenomenon that gave Stonebraker and DeWitt such heartburn~\cite{dewitt2008mapreduce}---are a re-realization of the Postgres idea of user-defined code hosted in a query framework. MapReduce looks very much like a combination of software engineering ideas from Postgres combined with parallelism ideas from systems like Gamma and Teradata, with some minor innovation around mid-query restart for extreme-scalability workloads. Postgres-based start-ups Greenplum and Aster showed around 2007 that parallelizing Postgres could result in something much higher-function and practical than MapReduce for most customers, but the market still wasn't ready for any of this technology in 2008. By now, in 2018, nearly every Big Data stack primarily serves a workload of parallel SQL with UDFs---very much like the design Stonebraker and team pioneered in Postgres.

\subsubsection{Extensible access methods for new data types}

Relational databases evolved around the same time as B-trees in the early 1970s, and B-trees helped fuel Codd's dream of "physical data independence": B-tree indexes provide a level of indirection that adaptively reorganizes physical storage without requiring applications to change. The main limitation of B-trees and related structures was that they only support equality lookups and 1-dimensional range queries. What if you have 2-dimensional range queries of the kind typical in mapping and CAD applications? This problem was \textit{au courant} at the time of Postgres, and the R-tree~\cite{Guttman:1984:RDI:602259.602266} developed by Antonin Guttman in Stonebraker's group was one of the most successful new indexes developed to solve this problem in practice. Still, the invention of an index structure does not solve the end-to-end systems problem of DBMS support for multi-dimensional range queries. Many questions arise. Can you add an access method like R-trees to your DBMS easily? Can you teach your optimizer that said access method will be useful for certain queries? Can you get concurrency and recovery correct?

This was a very ambitious aspect of the Postgres agenda: a software architecture problem affecting most of a database engine, from the optimizer to the storage layer and the logging and recovery system. R-trees became a powerful driver and the main example of the elegant extensibility of Postgres' access method layer and its integration into the query optimizer. Postgres demonstrated---in an opaque ADT style---how to register an abstractly described access method (the R-tree, in this case), and how a query optimizer could recognize an abstract selection predicate (a range selection in this case) and match it to that abstractly described access method. Questions of concurrency control were less of a focus in the original effort: the lack of a unidimensional ordering on keys made B-tree-style locking inapplicable.\footnote{The Postgres challenge of extensible access methods inspired one of my first research projects at the end of graduate school: the Generalized Search Trees (GiST)~\cite{Hellerstein:1995:GST:645921.673145} and subsequent notion of Indexability theory~\cite{Hellerstein:2002:MIB:505241.505244}. I implemented GiST in Postgres during a postdoc semester, which made it even easier to add new indexing logic in Postgres. Marcel Kornacker's thesis at Berkeley solved the difficult concurrency and recovery problems raised by extensible indexing in GiST in a templated way~\cite{Kornacker:1997:CRG:253260.253272}.}

PostgreSQL today leverages both the original software architecture of extensible access methods (it has B-tree, GiST, SP-GiST, and Gin indexes) and the extensibility and high concurrency of the Generalized Search Tree (GiST) interface as well. GiST indexes power the popular PostgreSQL-based PostGIS geographic information system; Gin indexes power PostgreSQL's internal text indexing support.

\subsubsection{Optimizer handling of queries with expensive UDFs}

In traditional query optimization, the challenge was generally to minimize the amount of tuple-flow (and hence I/O) you generate in processing a query. This meant that operators that filter tuples (selections) are good to do early in the query plan, while operators that can generate new tuples (join) should be done later. As a result, query optimizers would "push" selections below joins and order them arbitrarily, focusing instead on cleverly optimizing joins and disk accesses. UDFs changed this: if you have expensive UDFs in your selections, the order of executing UDFs can be critical to optimizing performance. Moreover, if a UDF in a selection is really time-consuming, it's possible that it should happen \textit{after} joins (i.e., selection "pullup"). Doing this optimally complicated the optimizer space.

I took on this problem as my first challenge in graduate school and it ended up being the subject of both my M.S. with Stonebraker at Berkeley and my Ph.D. at Wisconsin under Jeff Naughton, with ongoing input from Stonebraker. Postgres was the first DBMS to capture the costs and selectivities of UDFs in the database catalog. We approached the optimization problem by coming up with an optimal ordering of selections, and then an optimal interleaving of the selections along the branches of each join tree considered during plan search. This allowed for an optimizer that maintained the textbook dynamic programming architecture of System R, with a small additional sorting cost to get the expensive selections ordered properly.\footnote{When I started grad school, this was one of three topics that Stonebraker wrote on the board in his office as options for me to think about for a Ph.D. topic. I think the second was function indexing, but I cannot remember the third. }

The expensive function optimization feature was disabled in the PostgreSQL source trees early on, in large part because there weren't compelling use cases at that time for expensive user-defined functions.\footnote{Ironically, my code from grad school was fully deleted from the PostgreSQL source tree by a young open-source hacker named Neil Conway, who some years later started a Ph.D. with me at UC Berkeley and is now one of Stonebraker's Ph.D. grandchildren.} The examples we used revolved around image processing, and are finally becoming mainstream data processing tasks in 2018. Of course, today in the era of Big Data and machine learning workloads, expensive functions have become quite common, and I expect this problem to return to the fore. Once again, Postgres was well ahead of its time.

\subsection{Active Databases and Rule Systems}

The Postgres project began at the tail end of the AI community's interest in rule-based programming as a way to represent knowledge in ``expert systems.'' That line of thinking was not successful; many say it led to the much discussed "AI winter" that persisted through the 1990s. 

However, rule programming persisted in the database community in two forms. The first was theoretical work around declarative logic programming using Datalog. This was a bugbear of Stonebraker's; he really seemed to hate the topic and famously criticized it in multiple "community" reports over the years.\footnote{Datalog survived as a mathematical foundation for declarative languages, and has found application over time in multiple areas of computing including software-defined networks and compilers. Datalog is declarative querying ``on steroids'' as a fully expressive programming model. I was eventually drawn into it as a natural design choice, and have pursued it in a variety of applied settings outside of traditional database systems.} The second database rules agenda was pragmatic work on what was eventually dubbed Active Databases and Database Triggers, which evolved to be a standard feature of relational databases. Stonebraker characteristically voted with his feet to work on the more pragmatic variant.

\begin{lrbox}{\LstBox}
\begin{lstlisting}[
				   basicstyle={\tiny\ttfamily},
                   emph={int,char,double,float,unsigned},
                  ]
* DESCRIPTION:
* Take a deeeeeeep breath & read. If you can avoid hacking the code
* below (i.e. if you have not been "volunteered" by the boss to do this
* dirty job) avoid it at all costs. Try to do something less dangerous
* for your (mental) health. Go home and watch horror movies on TV.
* Read some Lovecraft. Join the Army. Go and spend a few nights in
* people's park. Commit suicide...
* Hm, you keep reading, eh? Oh, well, then you deserve what you get.
* Welcome to the gloomy labyrinth of the tuple level rule system, my
* poor hacker...
\end{lstlisting}
\end{lrbox}

Stonebraker's work on database rules began with Eric Hanson's Ph.D., which initially targeted Ingres but quickly transitioned to the new Postgres project. It expanded to the Ph.D. work of Spyros Potamianos on PRS2: Postgres Rules System 2. A theme in both implementations was the potential to implement rules in two different ways. One option was to treat rules as query rewrites, reminiscent of the work on rewriting views that Stonebraker pioneered in Ingres. In this scenario, a rule logic of "on condition then action" is recast as "on query then rewrite to a modified query and execute it instead." For example, a query like "append a new row to Mike's list of awards" might be rewritten as "raise Mike's salary by 10\%." The other option was to implement a more physical "on condition then action," checking conditions at a row level by using locks inside the database. When such locks were encountered, the result was not to wait (as in traditional concurrency control), but to execute the associated action.\footnote{The code for row-level rules in PRS2 was notoriously tricky. A bit of searching in the Berkeley Postgres archives unearthed the following source code comment---probably from Spyros Potamianos---in Postgres version 3.1, circa 1991: \usebox{\LstBox}
}

In the end, neither the query rewriting scheme nor the row-level locking scheme was declared a "winner" for implementing rules in Postgres---both were kept in the released system. Eventually all of the rules code was scrapped and rewritten in PostgreSQL, but the current source still retains both the notions of per-statement and per-row triggers.

The Postgres rules systems were very influential in their day, and went "head-to-head" with research from IBM's Starburst project and MCC's HiPac project. Today, "triggers" are part of the SQL standard and implemented in many of the major database engines. They are used somewhat sparingly, however. One problem is that this body of work never overcame the issues that led to AI winter: the interactions within a pile of rules can become untenably confusing as the rule set grows even modestly. In addition, triggers still tend to be relatively time-consuming in practice, so database installations that have to run fast tend to avoid the use of triggers. But there has been a cottage industry in related areas like materialized view maintenance, Complex Event Processing and stream queries, all of which are in some way extensions of ideas explored in the Postgres rules systems.

\subsection{Log-centric Storage and Recovery}

Stonebraker described his design for the Postgres storage system this way:

\begin{quote}When considering the POSTGRES storage system, we were guided by a missionary zeal to do something different. All current commercial systems use a storage manager with a write-ahead log (WAL), and we felt that this technology was well understood. Moreover, the original Ingres prototype from the 1970s used a similar storage manager, and we had no desire to do another implementation.~\cite{Stonebraker:1991:PNG:125223.125262}
\end{quote}

While this is cast as pure intellectual restlessness, there were technical motivations for the work as well. Over the years, Stonebraker repeatedly expressed distaste for the complex write-ahead logging schemes pioneered at IBM and Tandem for database recovery. One of his core objections was based on a software engineering intuition that nobody should rely upon something that complicated---especially for functionality that would only be exercised in rare, critical scenarios after a crash.

The Postgres storage engine unified the notion of primary storage and historical logging into a single, simple disk-based representation. At base, the idea was to keep each record in the database in a linked list of versions stamped with transaction IDs---in some sense, this is ``the log as data'' or ``the data as a log,'' depending on your point of view. The only additional metadata required is a list of committed transaction IDs and wall-clock times. This approach simplifies recovery enormously, since there's no ``translating'' from a log representation back to a primary representation. It also enables ``time-travel'' queries: you can run queries ``as of'' some wall-clock time, and access the versions of the data that were committed at that time. The original design of the Postgres storage system---which reads very much as if Stonebraker wrote it in one creative session of brainstorming---contemplated a number of efficiency problems and optimizations to this basic scheme, along with some wet-finger analyses of how performance might play out~\cite{Stonebraker:1987:DPS:645914.671639}. The resulting implementation in Postgres was somewhat simpler.

Stonebraker's idea of ``radical simplicity'' for transactional storage was deeply counter-cultural at the time, when the database vendors were differentiating themselves by investing heavily in the machinery of high-performance transaction processing. Benchmark winners at the time achieved high performance and recoverability via highly optimized, complex write-ahead logging systems. Once they had write-ahead logs working well, the vendors also began to innovate on follow-on ideas such as transactional replication based on log shipping, which would be difficult in the Postgres scheme. In the end, the Postgres storage system never excelled on performance; versioning and time-travel were removed from PostgreSQL over time and replaced by write-ahead logging.\footnote{Unfortunately, PostgreSQL still isn't particularly fast for transaction processing: its embrace of write-ahead logging is somewhat half-hearted. Oddly, the PostgreSQL team kept much of the storage overhead of Postgres tuples to provide multiversion \textit{concurrency control}, something that was never a goal of the Berkeley Postgres project. The result is a storage system that can emulate Oracle's snapshot isolation with a fair bit of extra I/O overhead, but one that does not support Stonebraker's original idea of time travel or simple recovery.

Mike Olson notes that his original intention was to replace the Postgres B-tree implementation with his own B-tree implementation from the BerkeleyDB project, which developed at Berkeley during the Postgres era. But Olson never found the time. When Berkeley DB got transactional support years later at Sleepycat Corp., Olson tried to persuade the (then-) PostgreSQL community to adopt it for recovery, in place of no-overwrite. They declined; there was a hacker on the project who desperately wanted to build an MVCC system, and as that hacker was willing to do the work, he won the argument.

Although the PostgreSQL storage engine is slow, that is not intrinsic to the system. The Greenplum fork of PostgreSQL integrated an interesting alternative high-performance compressed storage engine. It was designed by Matt McCline---a veteran of Jim Gray's team at Tandem. It also did not support time travel.} But the time-travel functionality was interesting and remained unique. Moreover, Stonebraker's ethos regarding simple software engineering for recovery has echoes today both in the context of NoSQL systems (which choose replication rather than write-ahead logging) and main-memory databases (which often use multi-versioning and compressed commit logs).  The idea of versioned relational databases and time-travel queries are still relegated to esoterica today, popping up in occasional research prototypes and minor open-source projects. It is an idea that is ripe for a comeback in our era of cheap storage and continuously streaming data.

\subsection{Queries over New Deep Storage Technologies}

In the middle of the Postgres project, Stonebraker signed on as a co-PI on a large grant for digital earth science called Project Sequoia. Part of the grant proposal was to handle unprecedented volumes of digital satellite imagery requiring up to 100 terabytes of storage, far more data than could be reasonably stored on magnetic disks at the time. The center of the proposed solution was to explore the idea of a DBMS (namely Postgres) facilitating access to near-line ``tertiary'' storage provided by robotic ``jukeboxes'' for managing libraries of optical disks or tapes. 

A couple different research efforts came out of this. One was the Inversion file system: an effort to provide a UNIX filesystem abstraction \textit{above} an RDBMS. In an overview paper for Sequoia, Stonebraker described this in his usual cavalier style as ``a straightforward exercise''~\cite{stonebraker1995overview}. In practice, this kept Stonebraker student (and subsequent Cloudera founder) Mike Olson busy for a couple years, and the final result was not exactly straightforward~\cite{inversion} nor did it survive in practice.\footnote{Some years after Inversion, Bill Gates tilted against this same windmill with WinFS, an effort to rebuild the most widely-used filesystem in the world over a relational database backend. WinFS was delivered in developer releases of Windows but never made it to market. Gates later called this his greatest disappointment at Microsoft.}

The other main research thrust on this front was the incorporation of tertiary storage into a more typical relational database stack, which was the subject of Sunita Sarawagi's Ph.D. thesis. The main theme was to change the scale at which you think about managing space (i.e., data in storage and the memory hierarchy) and time (coordinating query and cache scheduling to minimize undesirable I/Os). One of the key issues in that work was to store and retrieve large multidimensional arrays in tertiary storage---echoing work in multidimensional indexing, the basic ideas included breaking up the array into chunks, and storing chunks together that are fetched together---including replicating chunks to enable multiple physical ``neighbors'' for a given chunk of data. A second issue was to think about how disk becomes a cache for tertiary storage. Finally, query optimization and scheduling had to take into account the long load times of tertiary storage and the importance of ``hits'' in the disk cache---this affects both the plan chosen by a query optimizer, and the time at which that plan is scheduled for execution.

Tape and optical disk robots are not widely used at present. But the issues of tertiary storage are very prevalent in the cloud, which has deep storage hierarchies in 2018: from attached solid-state disks to reliable disk-like storage services (e.g., AWS EBS) to archival storage (e.g., AWS S3) to deep storage (e.g., AWS Glacier). It is still the case today that these storage tiers are relatively detached, and there is little database support for reasoning about storage across these tiers. I would not be surprised if the issues explored on this front in Postgres are revisited in the near term.

\subsection{Support for Multiprocessors: XPRS}

Stonebraker never architected a large parallel database system, but he led many of the motivating discussions in the field. His ``Case for Shared Nothing'' paper~\cite{stonebraker1986case} documented the coarse-grained architectural choices in the area; it popularized the terminology used by the industry, and threw support behind shared-nothing architectures like those of Gamma and Teradata, which were rediscovered by the Big Data crowd in the 2000s.

Ironically, Stonebraker's most substantive contribution to the area of parallel databases was a ``shared-memory'' architecture called XPRS, which stood for eXtended Postgres on RAID and Sprite. XPRS was the ``Justice League'' of Berkeley systems in the early 1990s: a brief combination of Stonebraker's Postgres system, John Ousterhout's Sprite distributed OS, and Dave Patterson's and Randy Katz's RAID storage architectures. Like many multi-faculty efforts, the execution of XPRS was actually determined by the grad students who worked on it. The primary contributor ended up being Wei Hong, who wrote his Ph.D. thesis on parallel query optimization in XPRS. Hence the main contribution of XPRS to the literature and industry was parallel query optimization, with no real consideration of issues involving RAID or Sprite.\footnote{Of the three projects, Postgres and RAID both had enormous impact. Sprite is best remembered for Mendel Rosenblum's Ph.D. thesis on Log Structured File Systems (LFS), which had nothing of note to do with distributed operating systems. All three projects involved new ideas for disk storage beyond mutating single copies in place. LFS and the Postgres storage manager are rather similar, both rethinking logs as primary storage, and requiring expensive background reorganization. I once gently probed Stonebraker about rivalries or academic scoops between LFS and Postgres, but I never got any good stories out of him. Maybe it was something in the water in Berkeley at the time.}

In principle, parallelism ``blows up'' the plan space for a query optimizer by making it multiply the traditional choices made during query optimization (data access, join algorithms, join orders) against all possible ways of parallelizing each choice. The basic idea of what Stonebraker called ``The Wei Hong Optimizer'' was to cut the problem in two: run a traditional single-node query optimizer in the style of System R, and then ``parallelize'' the resulting single-node query plan by scheduling the degree of parallelism and placement of each operator based on data layouts and system configuration. This approach is heuristic, but it makes parallelism an additive cost to traditional query optimization, rather than a multiplicative cost. 

Although ``The Wei Hong Optimizer'' was designed in the context of Postgres, it became the standard approach for many of the parallel query optimizers in industry.

\subsection{Support for a Variety of Language Models}

One of Stonebraker's recurring interests since the days of Ingres was the programmer API to a database system. In his Readings in Database Systems series, he frequently included work like Carlo Zaniolo's GEM language as important topics for database system aficionados to understand. This interest in language undoubtedly led him to partner up with Larry Rowe on Postgres, which in turn deeply influenced the design of the Postgres data model and its Object-Relational approach. Their work focused largely on data-centric applications they saw in the commercial realm, including both business processing and emerging applications like CAD/CAM and GIS. 

One issue that was forced upon Stonebraker at the time was the idea of ``hiding'' the boundary between programming language constructs and database storage. Various competing research projects and companies exploring Object-Oriented Databases (OODBs) were targeting the so-called ``impedance mismatch'' between imperative object-oriented programming languages like Smalltalk, C++, and Java, and the declarative relational model. The OODB idea was to make programming language objects be optionally marked ``persistent,'' and handled automatically by an embedded DBMS. Postgres supported storing nested objects and ADTs, but its relational-style declarative query interface meant that each roundtrip to the database was unnatural for the programmer (requiring a shift to declarative queries) and expensive to execute (requiring query parsing and optimization). To compete with the OODB vendors, Postgres exposed a so-called ``Fast Path'' interface: basically a C/C++ API to the storage internals of the database. This enabled Postgres to be moderately performant in academic OODB benchmarks, but never really addressed the challenge of allowing programmers in multiple languages to avoid the impedance mismatch problem. Instead, Stonebraker branded the Postgres model as ``Object-Relational,'' and simply sidestepped the OODB workloads as a ``zero-billion dollar'' market.  Today, essentially all commercial relational database systems are ``Object-Relational'' database systems.

This proved to be a sensible decision. Today, none of the OODB products exist in their envisioned form, and the idea of ``persistent objects'' in programming languages has largely been discarded. By contrast, there is widespread usage of object-relational mapping layers (fueled by early efforts like Java Hibernate and Ruby on Rails) that allow declarative databases to be tucked under nearly any imperative object-oriented programming language as a library, in a relatively seamless way. This application-level approach is different than both OODBs and Stonebraker's definition of Object-Relational DBs.  In addition, lightweight persistent key-value stores have succeeded as well, in both non-transactional and transactional forms. These were pioneered by Stonebraker's Ph.D. student Margo Seltzer, who wrote BerkeleyDB as part of her Ph.D. thesis at the same time as the Postgres group, which presaged the rise of distributed ``NoSQL'' key-value stores like Dynamo, MongoDB, and Cassandra.

\section{Software Impact}

\subsection{Open Source}

Postgres was always an open source project with steady releases, but in its first many years it was targeted at usage in research, not in production.  

As the Postgres research project was winding down, two students in Stonebraker's group---Andrew Yu and Jolly Chen---modified the system's parser to accept an extensible variant of SQL rather than the original Postquel language. The first Postgres release supporting SQL was Postgres95; the next was dubbed PostgreSQL.

A set of open-source developers became interested in PostgreSQL and ``adopted'' it even as the rest of the Berkeley team was moving on to other interests. Over time the core developers for PostgreSQL have remained fairly stable, and the open-source project has matured enormously. Early efforts focused on code stability and user-facing features, but over time the open source community made significant modifications and improvements to the core of the system as well, from the optimizer to the access methods and the core transaction and storage system. Since the mid-1990s, very few of the PostgreSQL internals came out of the academic group at Berkeley---the last contribution may have been my GiST implementation in the latter half of the 1990s---but even that was rewritten and cleaned up substantially by open-source volunteers (from Russia, in that case). The open source community around PostgreSQL deserves enormous credit for running a disciplined process that has soldiered on over decades to produce a remarkably high-impact and long-running project. 

While many things have changed in 25 years, the basic architecture of PostgreSQL remains quite similar to the university releases of Postgres in the early 1990s, and developers familiar with the current PostgreSQL source code would have little trouble wandering through the Postgres3.1 source code (c. 1991). Everything from source code directory structures to process structures to data structures remain remarkably similar. The code from the Berkeley Postgres team had excellent bones.

PostgreSQL today is without question the most high-function open-source DBMS, supporting features that are often missing from commercial products. It is also (according to one influential rankings site) the most popular widely used independent open source database in the world\footnote{According to DB Engines, PostgreSQL today is the fourth most popular DBMS in the world, after Oracle, MySQL and MS SQL Server, all of which are corporate offerings (MySQL was acquired by Oracle many years ago)~\cite{dbengines}.  See the DB-Engines ranking methodology for a discussion of the rules for this ranking~\cite{dbengines-methodology}.} and its impact continues to grow: in both 2017 and 2018 it was the fastest-growing database system in the world in popularity~\cite{dbengines-dbms-of-the-year} PostgreSQL is used across a wide variety of industries and applications, which is perhaps not surprising given its ambition of broad functionality.

Heroku is a cloud SaaS provider that is now part of Salesforce. Postgres was adopted by Heroku in 2010 as the default database for its platform. Heroku chose Postgres because of its operational reliability. With Heroku's support, more major application frameworks such as Ruby on Rails and Python for Django began to recommend Postgres as their default database. 

PostgreSQL today supports an extension framework that makes it easy to add additional functionality to the system via UDFs and related modifications. There is now an ecosystem of PostgreSQL extensions---akin to the Illustra vision of DataBlades, but in open source. Some of the more interesting extensions include the Apache MADlib library for machine learning in SQL, and the Citus library for parallel query execution.

One of the most interesting open-source applications built over Postgres is the PostGIS Geographic Information System, which takes advantage of many of the features in Postgres that originally inspired Stonebraker to start the project.

\subsection{Commercial Adaptations}

PostgreSQL has long been an attractive starting point for building commercial database systems, given its permissive open source license, its robust codebase, its flexibility, and breadth of functionality. Summing the acquisition prices listed below, Postgres has led to over \$2.6 billion in acquisitions.\footnote{Note that this is a measure in real transaction dollars, and is much more substantial than the values often thrown around in high tech. Numbers in the billions are often used to describe estimated value of stock holdings, but are often inflated by 10x or more against contemporary value in hopes of future value. The transaction dollars of an acquisition measure the actual market value of the company at the time of acquisition. It is fair to say that Postgres has generated more than \$2.6 billion of real commercial value.} Many of the commercial efforts that built on PostgreSQL have addressed what is probably its key limitation: the ability to scale out to a parallel, shared-nothing architecture.\footnote{Parallelizing PostgreSQL requires a fair bit of work, but is eminently doable by a small, experienced team. Today, industry-managed open-source forks of PostgreSQL such as Greenplum and CitusDB offer this functionality. It is a shame that PostgreSQL wasn't parallelized in a true open source way much earlier. If PostgreSQL had been extended with shared-nothing features in open source in the early 2000s, it is quite possible that the open source Big Data movement would have evolved quite differently and more effectively.}

\begin{enumerate}[1. ]

\item Illustra was Stonebraker's second major start-up company, founded in 1992, seeking to commercialize Postgres as RTI had commercialized Ingres.\footnote{Illustra was actually the third name proposed for the company. Following the painterly theme established by Ingres, Illustra was originally called Mir\'{o}. For trademark reasons the name was changed to Montage, but that also ran into trademark problems.} The founding team included some of the core Postgres team including recent Ph.D. alumnus Wei Hong and then-chief programmer Jeff Meredith, along with Ingres alumni Paula Hawthorn and Michael Ubell. Postgres M.S. student Mike Olson joined shortly after the founding, and I worked on the Illustra handling of optimizing expensive functions as part of my Ph.D. work. There were three main efforts in Illustra: to extend SQL92 to support user-defined types and functions as in Postquel, to make the Postgres code base robust enough for commercial use, and to foster the market for extensible database servers via examples of ``DataBlades,'' domain-specific plug-in components of data types and functions. Illustra was acquired by Informix in 1997 for an estimated \$400M~\cite{illustra} and its DataBlade architecture was integrated into a more mature Informix query processing codebase as Informix Universal Server.

\item Netezza was a startup founded in 1999, which forked the PostgreSQL codebase to build a high-performance parallel query processing engine on custom FPGA-based hardware. Netezza was quite successful as an independent company, and had its IPO in 2007. It was eventually acquired by IBM, with a value of \$1.7B~\cite{netezza}.

\item Greenplum was the first effort to offer a shared-nothing parallel, scale-out version of PostgreSQL. Founded in 2003, Greenplum forked from the public PostgreSQL distribution, but maintained the APIs of PostgreSQL to a large degree, including the APIs for user-defined functions. In addition to parallelization, Greenplum extended PostgreSQL with an alternative high-performance compressed columnar storage engine, and a parallelized rule-driven query optimizer called Orca. Greenplum was acquired by EMC in 2010 for an estimated \$300M~\cite{greenplum}; in 2012, EMC consolidated Greenplum into its subsidiary, Pivotal. In 2015, Pivotal chose to release Greenplum and Orca back into open source. One of the efforts at Greenplum that leveraged its Postgres API was the MADlib library for machine learning in SQL~\cite{hellerstein2012madlib}. MADlib lives on today as an Apache project. Another interesting open-source project based on Greenplum is Apache HAWQ, a Pivotal design that runs the ``top half'' of Greenplum (i.e., the parallelized PostgreSQL query processor and extensibility APIs) in a decoupled fashion over Big Data stores such as the Hadoop File System.

\item EnterpriseDB was founded in 2004 as an open-source-based business, selling PostgreSQL in both a vanilla and enhanced edition with related services for enterprise customers. A key feature of the enhanced EnterpriseDB Advanced Server is a set of database compatibility features with Oracle, to allow application migration off of Oracle.

\item Aster Data was founded in 2005 by two Stanford students, to build a parallel engine for analytics. Its core single-node engine was based on PostgreSQL. Aster focused on queries for graphs and on analytics packages based on UDFs that could be programmed with either SQL or MapReduce interfaces. Aster Data was acquired by Teradata in 2011 for \$263M~\cite{aster}. While Teradata never integrated Aster into its core parallel database engine, it still maintains Aster as a standalone product for use cases outside the core of Teradata's warehousing market.

\item ParAccel was founded in 2006, selling a shared-nothing parallel version of PostgreSQL with column-oriented, shared-nothing storage. ParAccel enhanced the Postgres optimizer with new heuristics for queries with many joins. In 2011, Amazon invested in ParAccel, and in 2012 announced AWS Redshift, a hosted data warehouse as a service in the public cloud based on ParAccel technology. In 2013, ParAccel was acquired by Actian (who also had acquired Ingres) for an undisclosed amount---meaning it was not a material expense for Actian. Meanwhile, AWS Redshift has been an enormous success for Amazon---for many years it was the fastest-growing service on AWS, and many believe it is poised to put long-time data warehousing products like Teradata and Oracle Exadata out of business. In this sense, Postgres may achieve its ultimate dominance in the cloud.

\item CitusDB was founded in 2010 to offer a shared-nothing parallel implementation of PostgreSQL. While it started as a fork of PostgreSQL, as of 2016 CitusDB is implemented via public PostgreSQL extension APIs and can be installed into a vanilla PostgreSQL installation. Also as of 2016, the CitusDB extensions are available in open source. 

\end{enumerate}
\section{Lessons}

You can draw a host of lessons from the success of Postgres, a number of them defiant of conventional wisdom. 

The highest-order lesson I draw comes from the fact that that Postgres defied Fred Brooks' ``Second System Effect''~\cite{brooks1975mythical}. Brooks argued that designers often follow up on a successful first system with a second system that fails due to being overburdened with features and ideas. Postgres was Stonebraker's second system, and it was certainly chock full of features and ideas. Yet the system succeeded in prototyping many of the ideas, while delivering a software infrastructure that carried a number of the ideas to a successful conclusion. This was not an accident---at base, Postgres was \textit{designed for extensibility}, and that design was sound. With extensibility as an architectural core, it is possible to be creative and stop worrying so much about discipline: you can try many extensions and let the strong succeed. Done well, the ``second system'' is not doomed; it benefits from the confidence, pet projects, and ambitions developed during the first system. This is an early architectural lesson from the more ``server-oriented'' database school of software engineering, which defies conventional wisdom from the ``component-oriented'' operating systems school of software engineering.

Another lesson is that a broad focus---``one size fits many''---can be a winning approach for both research and practice. To coin some names, ``MIT Stonebraker'' made a lot of noise in the database world in the early 2000s that ``one size doesn't fit all.'' Under this banner he launched a flotilla of influential projects and startups, but none took on the scope of Postgres. It seems that ``Berkeley Stonebraker'' defies the later wisdom of ``MIT Stonebraker,'' and I have no issue with that.\footnote{As Emerson said, ``a foolish consistency is the hobgoblin of little minds''.}  Of course there's wisdom in the ``one size doesn't fit all'' motto (it's always possible to find modest markets for custom designs!), but the success of ``Berkeley Stonebraker's'' signature system---well beyond its original intents---demonstrates that a broad majority of database problems can be solved well with a good general-purpose architecture. Moreover, the design of that architecture is a technical challenge and accomplishment in its own right. In the end---as in most science and engineering debates---there isn't only one good way to do things. Both Stonebrakers have lessons to teach us. But at base, I'm still a fan of the broader agenda that ``Berkeley Stonebraker'' embraced.

A final lesson I take from Postgres is the unpredictable potential that can come from open-sourcing your research. In his Turing talk, Stonebraker speaks about the ``serendipity'' of PostgreSQL succeeding in open source, largely via people outside Stonebraker's own sphere. It's a wonderfully modest quote: 

\begin{quote}
[A] pick-up team of volunteers, none of whom have anything to do with me or Berkeley, have been shepherding that open source system ever since 1995. The system that you get off the web for Postgres comes from this pick-up team.  It is open source at its best and I want to just mention that I have nothing to do with that and that collection of folks we all owe a huge debt of gratitude to~\cite{stonebrakerturing}.
\end{quote}

I'm sure all of us who have written open source would love for that kind of ``serendipity'' to come our way.  But it's not all serendipity---the roots of that good luck were undoubtedly in the ambition, breadth and vision that Stonebraker had for the project, and the team he mentored to build the Postgres prototype. If there's a lesson there, it might be to ``do something important and set it free.'' It seems to me (to use a Stonebrakerism) that you can't skip either part of that lesson.

\section{Acknowledgments}

I'm indebted to my old Postgres buddies Wei Hong, Jeff Meredith, and Mike Olson for their remembrances and input, and to Craig Kerstiens for his input on modern-day PostgreSQL.

\bibliography{stonebraker.bib}

\end{document}